%% file: main.tex
\documentclass[twocolumn]{aastex631}

\newcommand\latex{La\TeX}

\newcommand{\nhim}{N_{\rm H}^{\rm IM}}
\newcommand{\nh}{N_{\rm H}}
\newcommand{\pcmcm}{\;{\rm cm^{-2}}}
\newcommand{\alfven}{Alfv\'{e}n}
\newcommand{\nhpe}{N_{\rm H}^{\rm PE}}
\newcommand{\nhcs}{N_{\rm H}^{\rm CS}}

\newcommand{\porb}{P_{\rm orb}}
\newcommand{\pspin}{P_{\rm spin}}
\newcommand{\axsini}{a_{\rm x}\sin i}

\usepackage{comment}
\usepackage{multirow}
\usepackage{enumerate}
\usepackage{mathtools}
\usepackage{amsmath}

\submitjournal{ApJ}

\shorttitle{Cen X-3 NuSTAR}
\shortauthors{Tamba et al.}
\graphicspath{{./}{figures/}}

\newcommand{\red}[1]{{#1}}

\defcitealias{Tamba2023}{Paper I}

\begin{document}

\title{The nature of spectral variability of the accreting pulsar Centaurus X-3 unveiled by NuSTAR observation covering two orbital cycles}

\correspondingauthor{Tsubasa Tamba, Hirokazu Odaka}
\email{tamba@ac.jaxa.jp odaka@ess.sci.osaka-u.ac.jp}

\author[0000-0001-7631-4362]{Tsubasa Tamba}
\affiliation{Japan Aerospace Exploration Agency, Institute of Space and Astronautical Science, 3-1-1 Yoshino-dai, Chuo-ku, Sagamihara, Kanagawa 252-5210, Japan}
\affiliation{Department of Physics, Faculty of Science, The University of Tokyo, 7-3-1 Hongo, Bunkyo-ku, Tokyo 113-0033, Japan}

\author[0000-0003-2670-6936]{Hirokazu Odaka}
\affiliation{Department of Earth and Space Science, Osaka University, 1-1 Machikaneyama-cho, Toyonaka, Osaka 560-0043, Japan}
\affiliation{Department of Physics, Faculty of Science, The University of Tokyo, 7-3-1 Hongo, Bunkyo-ku, Tokyo 113-0033, Japan}
\affiliation{Kavli IPMU, The University of Tokyo, Kashiwa 113-0033, Japan}
%\affiliation{Research Center for Early Universe, Faculty of Science, The University of Tokyo, 7-3-1 Hongo, Bunkyo-ku, Tokyo 113-0033, Japan}

\author[0000-0002-0114-5581]{Atsushi Tanimoto}
\affiliation{Graduate School of Science and Engineering, Kagoshima University, 1-21-24, Korimoto, Kagoshima, Kagoshima 890-0065, Japan}
%\affiliation{Department of Physics, Faculty of Science, The University of Tokyo, 7-3-1 Hongo, Bunkyo-ku, Tokyo 113-0033, Japan}

\author[0000-0003-4237-1101]{Hiromasa Suzuki}
\affiliation{Japan Aerospace Exploration Agency, Institute of Space and Astronautical Science, 3-1-1 Yoshino-dai, Chuo-ku, Sagamihara, Kanagawa 252-5210, Japan}

\author[0000-0003-0590-6330]{Satoshi Takashima}
\affiliation{Department of Physics, Faculty of Science, The University of Tokyo, 7-3-1 Hongo, Bunkyo-ku, Tokyo 113-0033, Japan}

\author[0000-0003-0890-4920]{Aya Bamba}
\affiliation{Department of Physics, Faculty of Science, The University of Tokyo, 7-3-1 Hongo, Bunkyo-ku, Tokyo 113-0033, Japan}
\affiliation{Research Center for Early Universe, Faculty of Science, The University of Tokyo, 7-3-1 Hongo, Bunkyo-ku, Tokyo 113-0033, Japan}
\affiliation{Trans-Scale Quantum Science Institute, The University of Tokyo, Tokyo  113-0033, Japan}

\begin{abstract}

\input{abstract.tex}

\end{abstract}

\input{section1.tex}

\input{section2.tex}

\input{section3.tex}

\input{section4.tex}

\input{section5.tex}

\input{section6.tex}

\input{section7.tex}

\begin{acknowledgments}

This work is supported by JSPS/MEXT KAKENHI grant Nos. 23KJ2214 (T.T.), 22H00128, 22K18277 (HO), 20J00119 (AT), JP19K03908, and JP23H01211 (A.B.).

\end{acknowledgments}

\vspace{5mm}
\facilities{NuSTAR \citep{Harrison2013}}


\software{HEASoft, Xronos, XSPEC \citep{Arnaud1996}, matplotlib \citep{Hunter2007}
          }

\bibliography{reference}{}
\bibliographystyle{aasjournal}


\end{document}

%% file: abstract.tex
We conducted a $369\;{\rm ks}$ NuSTAR observation on the X-ray pulsar Centaurus~X-3, which covered two consecutive orbital cycles of the source, including two eclipse durations.
We investigated the orbital-phase spectral variability over the two orbital cycles.
We divided the entire observation data into multiple segments, each covering an orbital interval of $\Delta\Phi=0.005$.
The phenomenological spectral modeling applied to these orbital-phase-resolved spectra reveals that the photon index is the key parameter with the most variability and a strong correlation with the continuum flux.
The photon index becomes softer during the high-flux phases and harder in the low-flux phases.
The relation between the photon index and continuum flux remains consistent when investigating specific spin phases, suggesting that the spectral variability originates from extrinsic factors apart from the neutron star.
Furthermore, the 3--5 keV pulse fraction also exhibits variability, being enhanced in the high-flux phases and suppressed in the low-flux phases, which indicates the presence of multiple emission components with different pulse fractions.
These phenomenological analysis results enabled us to estimate the physical origin of the spectral variability.
We successfully fitted the orbital-phase-resolved spectra with a physical model that assumes (1) stable emission from the neutron star, (2) attenuation by inhomogeneous, clumpy stellar wind, and (3) an additional non-pulsed emission component arising from thermal emission from the accretion disk.
The thermal emission from the accretion disk can be described by a blackbody with a temperature of $kT\sim0.5\;{\rm keV}$ and a luminosity of $\sim10^{37}\;{\rm erg\;s^{-1}}$.

%% file: section1.tex
\section{Introduction} \label{sec:intro}

X-ray pulsars accompanying strong magnetic fields of $B\sim10^{12}\;{\rm G}$ provide ideal laboratories for studying physical processes in such extreme environments \citep[][for a review]{Mushtukov2022}.
The energy release from the accreting matter is observed as bright X-ray emissions from the accretion columns onto the magnetic poles.
In high-luminosity X-ray pulsars ($L_{\rm x}\gtrsim10^{37}\;{\rm erg\;s^{-1}}$; \citealp{Mushtukov2015}), the radiation from the neutron star decelerates the accretion flow, leading to the formation of a radiation-dominated shock inside the accretion column.
The X-ray spectrum from these objects can be explained by bulk and thermal Comptonization within the column \citep{Becker2005, Becker2007, Odaka2014}.
However, the currently prevailing theory is based on a 1D model along the magnetic axis, and our understanding of the 3D aspects of the accretion column radiation, such as the column geometry, radiation anisotropy, and the rotation of the neutron star, remains limited.

Obtaining 3D information about the accretion column radiation requires continuous, long-term observations that can support sophisticated spin-phase-resolved analysis.
However, many X-ray pulsars appear to change their accretion column states on even shorter timescales.
This is because most X-ray pulsars fall into the category of ``wind-fed'' X-ray pulsars, where flux changes are attributed to the highly variable capture rate of the inhomogeneous stellar wind \citep[e.g., Vela~X-1:][]{Kreykenbohm2008, Odaka2013, Odaka2014}.
Therefore, it is necessary to focus on the disk-fed X-ray pulsars, which provide more stable accretion rates due to the direct supply of the accretion flow from the Roche lobe of the companion star.

Centaurus~X-3 (Cen~X-3) is an ideal source for investigating the 3D physical characteristics of the accretion column.
\red{It stands out as one of the few X-ray pulsars associated with accretion disks, as indicated by its high luminosity of $\sim5\times10^{37}\;{\rm erg\;s^{-1}}$ \citep{Suchy2008} at a distance of $6.4^{+1.0}_{-1.4}\;{\rm kpc}$ \citep{Arnason2021}, the variability in the optical light curve \citep{Tjemkes1986}, and quasiperiodic oscillations at $\sim40\;{\rm mHz}$ \citep{Takeshima1991, Raichur2008_QPO}.}
\red{The accretion stream is predominantly considered to be disk-fed, although the contribution from wind-fed accretion cannot be ignored \citep[e.g.,][]{Petterson1978, Wojdowski2001, Day1993, Klawin2023}.}
Its well-measured orbital parameters also contribute to gaining straightforward insights into the geometry of the accretion column.
In this binary system, a neutron star with a mass of $1.34^{+0.16}_{-0.14}M_{\odot}$ \citep{vanderMeer2007} orbits a giant O6–8 III counterpart, V779 Cen.
The companion star has a mass and radius of $20.5\pm0.7M_{\odot}$ and $12R_{\odot}$ \citep{Schreier1972, Krzeminski1974, Hutchings1979}, respectively.
The orbital period is $\sim2.08\;{\rm days}$ \citep{Bildsten1997}, with the eclipse accounting for $\sim20\%$ of the orbit.
This eclipse duration has enabled us to estimate the inclination angle at $70.2\pm2.7\;{\rm deg}$ \citep{Ash1999}.
The neutron star itself has a spin period of $\sim4.8\;{\rm s}$, and gradual spin-up has been observed \citep{Tsunemi1996}.
Furthermore, a recent observation using IXPE (Imaging X-ray Polarimetry Explorer) has revealed that the source exhibits $5.8\%$ polarization in the 2--8 keV band with a position angle of $49.6^\circ$ \citep{Tsygankov2022}.
This polarization data offers additional insights for constraining the 3D characteristics of the source.

Despite the apparent stability of the accretion flow, Cen~X-3 still exhibits complex spectral variability, presumably due to a combination of internal factors (mild but non-negligible accretion rate variability) and external factors (circumstellar medium).
Unlike other disk-fed pulsars, the absence of superorbital modulation in the long-term light curve makes it an even more challenging problem \citep[e.g.,][]{Raichur2008}.
To investigate the physical pictures of the accretion column through long-term spin-phase-resolved analysis, it is essential to understand the mechanism responsible for the spectral variability.
We have already analyzed archival NuSTAR observation data and identified two comparable factors contributing to the spectral variability: intrinsic flux variability of $\sim10\%$ and obscuration by the clumpy stellar wind \citep[][hereafter \citetalias{Tamba2023}]{Tamba2023}.
However, this data set covered only $\sim0.2$ cycles of Cen~X-3, limiting our ability to make conclusions about the long-term variability.
To address this, we conducted another observation of Cen~X-3 with a significantly longer duration---an observation lasting $369\;{\rm ks}$, equivalent to two orbital cycles of the system.
This data set was already studied by \cite{Dangal2024} to investigate properties of cyclotron lines, \red{and by \cite{Liu2024} to investigate temporal and spectral variations}.
In this paper, we present the analysis results of the observation data, fucusing on the nature of the spectral variability of Cen~X-3.
This paper is organized as follows.
Section \ref{sec:observation_and_data_reduction} provides a brief overview of the observation and data reduction process.
Sections \ref{sec:spectral_variability} and \ref{sec:spin_phase_resolved} present phenomenological analyses based on orbital- and spin-phase-resolved observation data, respectively.
Building upon these analyses, we construct a physically-motivated spectral model and apply it to the observation data in Section \ref{sec:advanced_model}.
Finally, Sections \ref{sec:discussion} and \ref{sec:conclusions} offer discussions and conclusions, respectively.

%chatgpt done 2023/10/2

%For Cen~X-3, there have been two controversial hypotheses that explain the complex orbital-phase variability.
% definition of "internal" and "external"
% paper 1 suggest inhomogeneous stellar wind is the key cause

%%%%%%%%%%%%%%%%%%%%%%%%%%%%%%%%%%%%%%%%%%%%%%%%%%%%%%%%%%%%%%%%%%%%%%%%%%%%%%%%%%%%%%%%
%   comments 
%%%%%%%%%%%%%%%%%%%%%%%%%%%%%%%%%%%%%%%%%%%%%%%%%%%%%%%%%%%%%%%%%%%%%%%%%%%%%%%%%%%%%%%%

\begin{comment}

\end{comment}

%% file: section2.tex
\section{Observation and data reduction} \label{sec:observation_and_data_reduction}

We conducted a NuSTAR observation on Cen~X-3 from 2022 January 12 to 16 with an elapsed time of $369\;{\rm ks}$ (ObsID: 30701019002), which corresponds to $\sim2.04$ cycles of the binary system.
We followed the general data reduction procedure of NuSTAR observation data using {\tt HEASoft 6.30}, including barycentric correction.
The source region was defined as a circle with a radius of $180''$ centered on the source, while the background region was defined as a rectangle in the off-source region.
After the data reduction, we obtained cleaned event data for FPMA and FPMB with net exposures of $189\;{\rm ks}$ each.
All the temporal and spectral analyses in this paper were performed using {\tt Xronos} 6.0 and {\tt XSPEC} 12.12.1.

After completing the general data reduction, we applied binary correction to the arrival times of the observation data.
Subsequently, we conducted the $Z_{n}^2$ test \citep{Buccheri1983, Brazier1994} with $n=2$, a method commonly used for pulse searches.
In the pulse search, we treated the following four parameters as free variables: the orbital period $\porb$, the spin period $\pspin$, the projected semi-major axis $\axsini$, and the orbital phase at the start time of the observation $\Phi_{\rm start}$.
We determined the best-fit values of these four parameters as
\begin{eqnarray}
\porb&=&2.089(3)\;{\rm days},\\
\pspin&=&4.796653(23)\;{\rm s},\\
\axsini&=&39.65(82)\;{\rm lt\mathchar`-s},\\
\Phi_{\rm start}&=&-0.164(2),
\end{eqnarray}
where errors indicate $1\sigma$ confidence levels.
We note that $\Phi=0$ corresponds to the mid-eclipse of Cen~X-3.
The observation spanned from $\Phi=-0.164$ to $\Phi=1.880$ with an orbital-phase interval of $\Delta\Phi=2.044$.

% chatGPT done 2023/09/27

% obs and data reduction 
% spin period etc

%% file: section3.tex
\section{Spectral variability over two orbits} \label{sec:spectral_variability}

Figure \ref{fig:lc_and_hr}(a) shows the 3--78 keV light curve over the two orbital cycles.
It prominently exhibits eclipses within the ranges of $-0.1<\Phi<0.1$ and $<0.9<\Phi<1.1$.
Additionally, there are ingresses (pre-eclipse dips) and egresses (post-eclipse dips), each lasting for $\Delta\Phi\sim0.05$.
In order to examine the spectral variability along the orbital phase, we segmented the entire observation data into multiple orbital intervals.
For the out-of-eclipse phases, each orbital interval had a duration of $\Delta\Phi=0.005$, equivalent to $\sim900\;{\rm s}$.
During the eclipse phases, we assigned a single orbital interval with a much longer duration of $\Delta\Phi=0.05$ due to the limited photon counts.
The entire observation dataset was divided into 337 orbital intervals, with 98 of them being excluded from subsequent analysis due to short exposure times of $<150\;{\rm s}$.
In this section, we adopted a phenomenological approach to track the spectral variability by analyzing light curves and energy spectra within the remaining 239 orbital intervals.
\red{We also note that the obtained spectral parameters are based solely on a phenomenological model and do not represent any physical interpretation of Cen~X-3, which will be addressed in Section \ref{sec:advanced_model}.}

\subsection{Light curves}

\begin{figure*}[htb!]
\centering
\includegraphics[width=\linewidth]{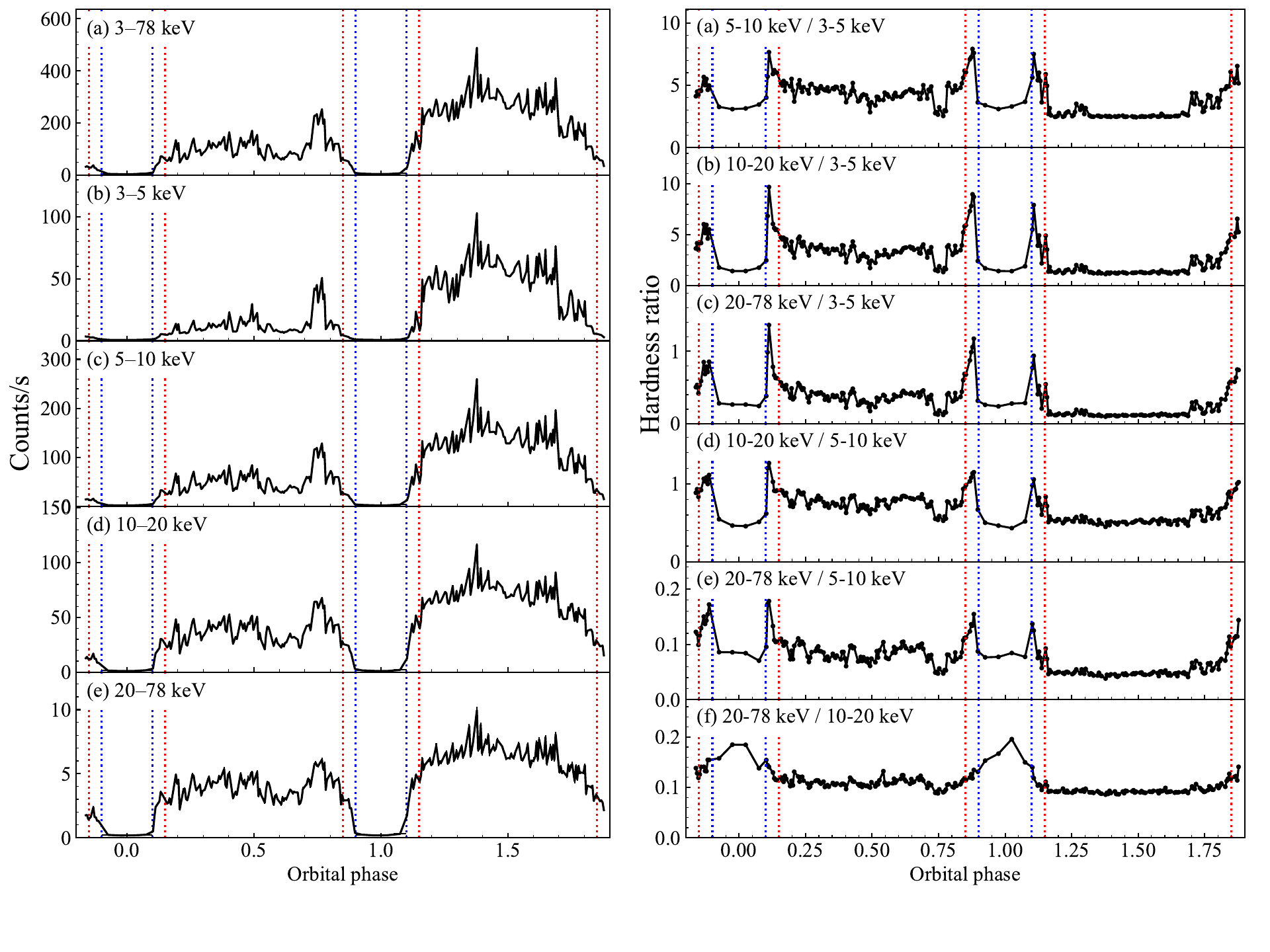}
\caption{Light curves for various energy bands (left) and hardness ratios among the energy bands (right).
The blue dotted lines indicate the start and end points of the eclipses.
The red lines represent the beginning of ingresses and the end of egresses.
}
\label{fig:lc_and_hr}
\end{figure*}

In order to monitor the spectral variability throughout the orbital phase, we first generated light curves for various energy bands.
The left panel of Figure \ref{fig:lc_and_hr} displays the light curves for four distinct energy ranges, 3--5, 5--10, 10--20, and 20--78 keV, along with their summation (3--78 keV).
All the energy bands exhibit noticeable variabilities along the orbital phase.
Interestingly, there is a consistent pattern where the second orbit (hereafter Orbit 2) exhibits higher count rates than the first orbit (hereafter Orbit 1).
The count ratios of Orbit 2 compared to Orbit 1 differ across the energy bands, with values of 3.62, 2.43, 1.78, and 1.55 for the 3--5, 5--10, 10--20, and 20--78 keV bands, respectively.
The gradual decrease in these ratios with increasing energy suggests that the lower energy bands exhibit higher variabilities along the orbital phase.

In the right panel of Figure \ref{fig:lc_and_hr}, we present the hardness ratios among the four energy bands as a function of the orbital phase.
Notably, the hardness ratios for Orbit 1 consistently exceed those for Orbit 2 across all panels.
A discernible pattern emerges, suggesting that orbital phases with higher flux levels exhibit ``softer'' spectra, while those with lower flux levels exhibit ``harder'' spectra.
It is worth noting that, with the exception of the 20--78 keV / 10--20 keV, the hardness ratios exhibit an increase during the ingress and egress phases, gradually returning to comparable values with Orbit 2 during the eclipse phases.

\subsection{Spectral analysis with phenomenological model}\label{subsec:phenomeno_spectroscopy}

We conducted spectral analysis on each spectrum generated from individual orbital intervals in order to investigate more details of the spectral variability.
We adopted the same phenomenological model as \citetalias{Tamba2023}.
The model is expressed as
\begin{eqnarray}
\frac{dN(E)}{dE}\propto f_{\rm phabs}\times\left(f_{\rm FDcut}+f_{\rm gauss}\right)\times f_{\rm gabs},
\label{eq:nu2_orb_phase_model}
\end{eqnarray}
where $f_{\rm phabs}$, $f_{\rm FDcut}$, $f_{\rm gauss}$, and $f_{\rm gabs}$ represent the photoelectric absorption by the interstellar medium and stellar wind \red{(cross section: \citealp{Verner1996}; solar abundance: \citealp{Anders1989})}, the continuum spectrum, Gaussian profile originating from the Fe line, and the absorption feature caused by CRSF (cyclotron resonance scattering feature), respectively.
The specific definitions of these functions can be found in \citetalias{Tamba2023}.
The hydrogen column density of the interstellar medium was held constant at $\nhim=1.1\times10^{22}\;{\rm cm^{-2}}$ \citep{HI4PI2016}.
\red{This constraint was introduced to align the lower bound of the total absorption with the ISM absorption and to prevent parameter coupling in the accurate evaluation of the stellar wind absorption.}
\red{Therefore, $f_{\rm phabs}$ was introduced as two separate multiplicative components, one of which is entirely fixed and one is set free.}
Given that the photon counts in individual orbital-phase-resolved spectra were insufficient to determine CRSF parameters accurately, we assumed that the CRSF component remained constant throughout the observation.
Our phase-averaged spectral analysis provided the following CRSF parameters: $E_{\rm cyc}=27.5\;{\rm keV}$, $\sigma_{\rm cyc}=6.31\;{\rm keV}$, and $\tau_{\rm cyc}=1.32$.
These values were fixed during the orbital-phase-resolved spectroscopy.
For the spectral fitting in this analysis, we considered the energy range of 4--78 keV.
The range of 3--4 keV was excluded \red{because the soft excess becomes dominant in this band \citep{Burderi2000}.}
We revisit this excluded band with a more sophisticated model, which we describe in Section \ref{sec:advanced_model}.

\begin{figure*}[htb!]
\centering
\includegraphics[width=\linewidth]{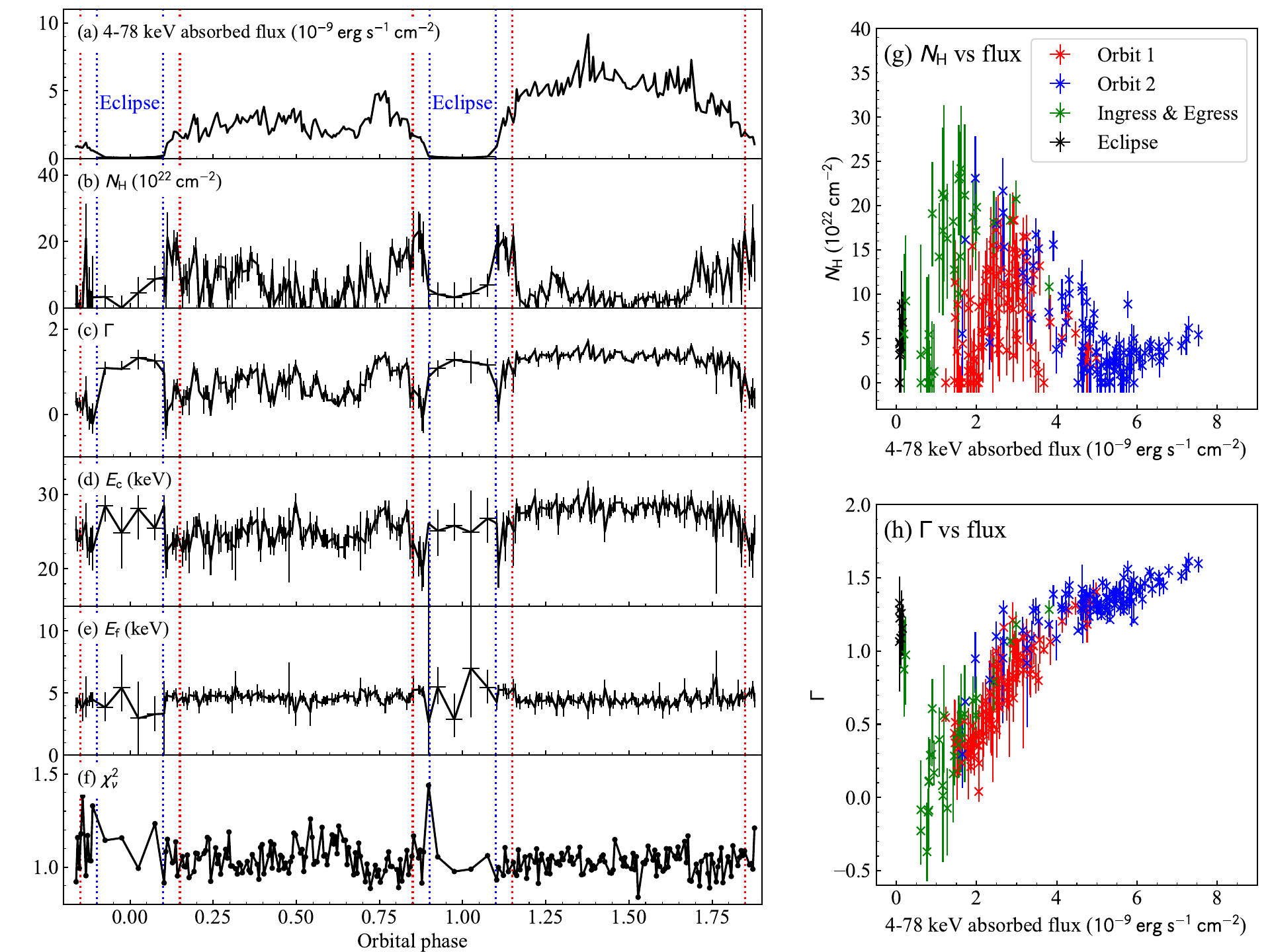}
\caption{(left) Orbital-phase variability of the best-fit spectral parameters derived from 4--78 keV spectroscopy.
(right) Correlation between $N_{\rm H}$ and flux (top), $\Gamma$ and flux (bottom). Different colors represent different types of orbital phases.
Error bars represent 90\% confidence levels.
}
\label{fig:orb_spec_params}
\end{figure*}

Figure \ref{fig:orb_spec_params} presents the outcomes of the orbital-phase-resolved spectroscopy.
All the spectra were successfully fitted by the model, and the spectral analysis yielded acceptable $\chi_{\nu}^2$ values, with an average of $1.04$.
The left panel shows the variability of the spectral parameters.
While the cut-off energy $E_{\rm c}$ and folding energy $E_{\rm f}$ remain relatively stable ($\chi_\nu^2=3.37$ and $0.60$ \red{with ${\rm d.o.f.}=238$} over the best-fit constant function, respectively), the hydrogen column density $N_{\rm H}$ and the photon index $\Gamma$ show significant variabilities along the orbital phase ($\chi_\nu^2=5.39$ and $20.31$ \red{with ${\rm d.o.f.}=238$}, respectively).
The right panels of Figure \ref{fig:orb_spec_params} illustrate the correlation between the spectral parameters responsible for the spectral variability ($\nh$ and $\Gamma$) and the continuum flux.
While $\nh$ does not exhibit any clear correlation with the continuum flux, $\Gamma$ shows a clear positive correlation with the flux.
Orbit 1 tends to have harder photon indices than Orbit 2, and during the ingress and egress phases, the photon indices become even harder.
Notably, the photon index returns to a soft value of $\sim1.2$ during the eclipse phases.

It becomes evident that the photon index $\Gamma$ is the crucial spectral parameter closely synchronized with the continuum flux.
The absorption column density $N_{\rm H}$ alone does not account for the flux variability.
This result suggests that the spectral variability cannot be solely attributed to varying levels of absorption arising from the inhomogeneous stellar wind.
Nonetheless, we have acquired two findings that imply the spectral variability originates from external factors.
Firstly, spectral hardening is observed during the ingress and egress phases \red{(Figure \ref{fig:orb_spec_params}c)}.
\red{Since the orbital motion should be independent of the accretion rate variability,} the presence of hard spectra during these phases is likely linked to the obscuration by the companion stellar wind.
The second finding pertains to the return of the photon index to softer values during the eclipse phases \red{(Figure \ref{fig:orb_spec_params}c)}.
The observed spectrum during the eclipse phases should closely resemble the original spectrum emitted from the neutron star, albeit with significantly reduced normalization.
This is because the continuum during the eclipse primarily results from Compton (and nearly Thomson) scattering by the surrounding stellar wind, preserving the energy distribution of the emitted photons.
The softening of the spectrum during the eclipse suggests that the original emission from the neutron star has a soft photon index of $\Gamma\sim1.2$, and it is hardened by external factors during the low-flux phases.

% chatgpt done 2023/09/27

%%%%%%%%%%%%%%%%%%%%%%%%%%%%%%%%%%%%%%%%%%%%%%%%%%%%%%%%%%%%%%%%%%%%%%%%%%%%%%%%%%%%%%%%
%   comments 
%%%%%%%%%%%%%%%%%%%%%%%%%%%%%%%%%%%%%%%%%%%%%%%%%%%%%%%%%%%%%%%%%%%%%%%%%%%%%%%%%%%%%%%%

\begin{comment}

\begin{figure}[htb!]
\centering
\includegraphics[width=240pt]{orbital_spectral_parameters_4-78keV.pdf}
\caption{Orbital-phase variation of spectral parameters derived from 4--78 keV orbital-phase-resolved spectra. The assumed model is Fermi-Dirac cut-off power-low with an Fe emission line (see text). Best-fit parameters with freed photon index (red) and fixed photon index of $\Gamma=1.21$ (blue) are plotted together. Error bars of fitting parameters represent 90\% confidence levels.}
\label{fig:orbital_spectral_parameters}
\end{figure}

\begin{figure*}[htb!]
\centering
\includegraphics[width=\linewidth]{flux_to_nh.pdf}
\caption{Relation between (a) 4--78 keV absorbed flux and $\nh$, (b) 4--78 keV unabsorbed flux and $\nh$, derived from orbital-phase-resolved spectroscopy. Best-fit parameters with freed photon index (red) and fixed photon index of $\Gamma=1.21$ (blue) are plotted together. Error bars of fitting parameters represent 90\% confidence levels.}
\label{fig:flux_to_nh}
\end{figure*}

\end{comment}

%% file: section4.tex
\section{Spin-phase-resolved analysis} \label{sec:spin_phase_resolved}

Examining the spin-phase variability is an additional valuable approach to nailing down the physical origin of the orbital-phase variability.
Our investigation concentrated on the spectral variabilities along both the orbital phase $\Phi$ and the spin phase $\phi$.
We aimed to uncover the connections between these two types of variability.
In this section, we present the results of our doubly-phase-resolved spectroscopy and pulse profile analysis.

\subsection{Doubly-phase-resolved spectroscopy}
\label{subsec:doubly_spectroscopy}

\begin{figure}[htb!]
\centering
\includegraphics[width=240pt]{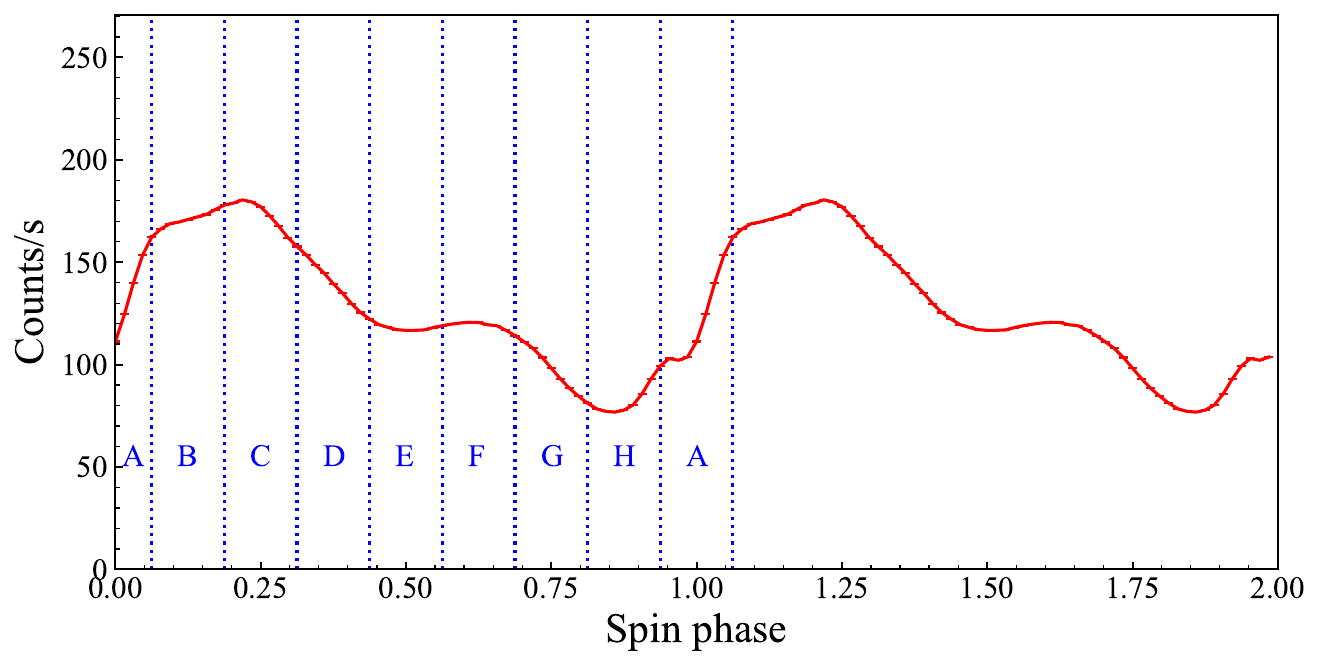}
\caption{\red{3--78 keV pulse profile with spin-phase intervals definition.}
}
\label{fig:spin_phase_intervals}
\end{figure}

We first generated a 3--78 keV pulse profile using the entire observation dataset \red{whose arrival times were corrected with the binary motion.}
\red{The pulse profile is shown in Figure \ref{fig:spin_phase_intervals}}.
The Fourier transform of this pulse profile enabled us to define the spin phase.
We designated $\phi=0$ as the starting point of the sinusoidal function and $\phi=0.25$ as the peak of the 1st harmonics.
Next, we divided the entire observation dataset into eight segments based on the spin phase.
These eight spin-phase intervals were denoted as A, B, ..., and H, each covering a spin-phase interval of $\Delta\phi=0.125$.
\red{This definition is depicted by blue dotted lines in Figure \ref{fig:spin_phase_intervals}.} 
Furthermore, we subdivided them into multiple segments based on the orbital phase, a procedure akin to what we conducted in Section \ref{subsec:phenomeno_spectroscopy}.
For each segment of the observation data, resolved by both orbital and spin phases, we conducted spectral analysis.

\begin{figure*}[htb!]
\centering
\includegraphics[width=\linewidth]{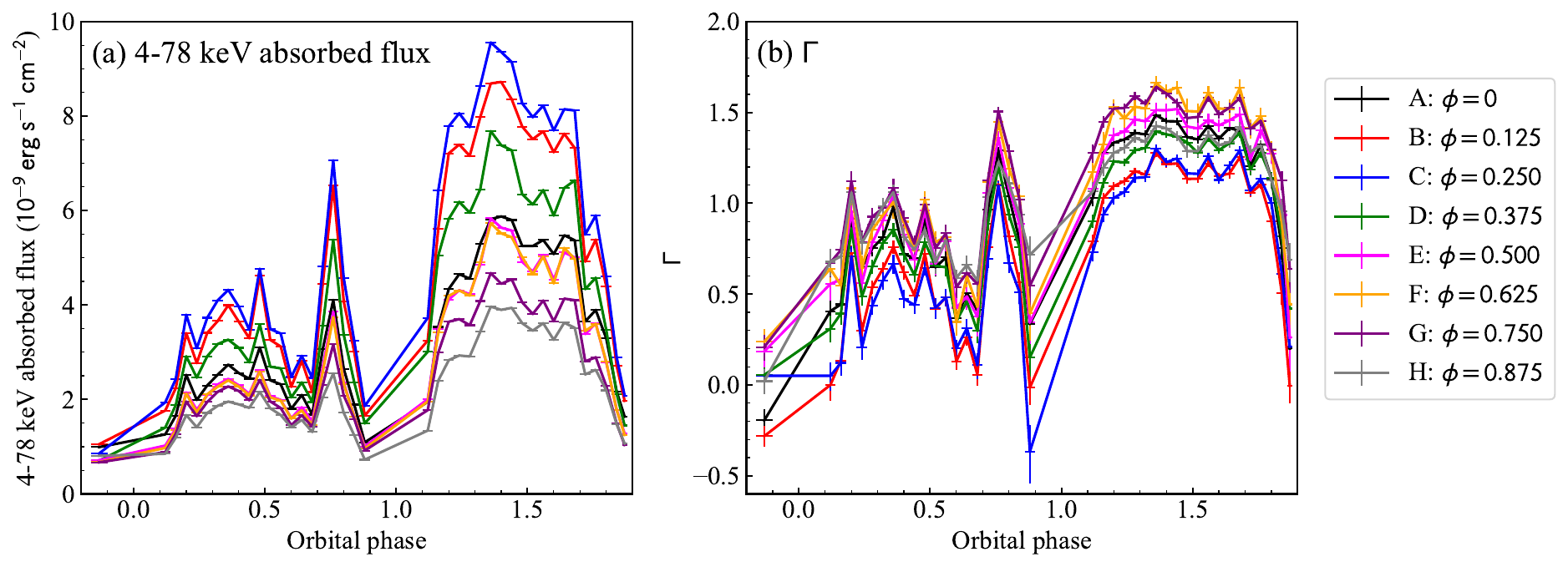}
\caption{Orbital-phase variabilities of 4--78 keV flux and photon index plotted for various spin phases.
Different colors correspond to distinct spin-phase intervals, labeled from A to H (see text).
Error bars represent 90\% confidence levels.
}
\label{fig:doubly_flux_and_gamma}
\end{figure*}

Figure \ref{fig:doubly_flux_and_gamma} presents the results of the doubly-phase-resolved spectroscopy.
The continuum flux (Figure \ref{fig:doubly_flux_and_gamma}a) exhibits very similar orbital-phase variations when considering specific spin-phase intervals.
This behavior closely resembles the trends observed in the spin-phase-averaged analysis (Figure \ref{fig:orb_spec_params}).
Likewise, the photon index (Figure \ref{fig:doubly_flux_and_gamma}b) displays parallel orbital-phase variabilities when examining particular spin-phase intervals, echoing the patterns observed in the spin-phase-averaged analysis (Figure \ref{fig:orb_spec_params}).
What is particularly noteworthy is that the relationships among the eight spin-phase intervals remain largely consistent throughout the two orbital cycles.
For instance, the spin-phase interval C (blue lines) consistently exhibits the highest flux and hardest photon index across all the orbital phases.
This result suggests that the variation in the photon index, which is predominantly synchronized with flux variability, is independent of the spin phase.
Instead, the flux variability appears to be driven by external factors originating separately from the neutron star.

%Both the continuum flux (Figure \ref{fig:doubly_flux_and_gamma}a) and the photon index (Figure \ref{fig:doubly_flux_and_gamma}b) display very similar trends of orbital-phase variabilities to those of spin-phase-averaged analysis (Figure \ref{fig:orb_spec_params}), even if we focus on a specific spin-phase interval.

%In this paper, we do not get into the physical origin that causes the spectral difference among these spin phases, which we will discuss elsewhere.

\subsection{Pulse profile}\label{subsec:pulse_profile}

\begin{figure}[htb!]
\centering
\includegraphics[width=240pt]{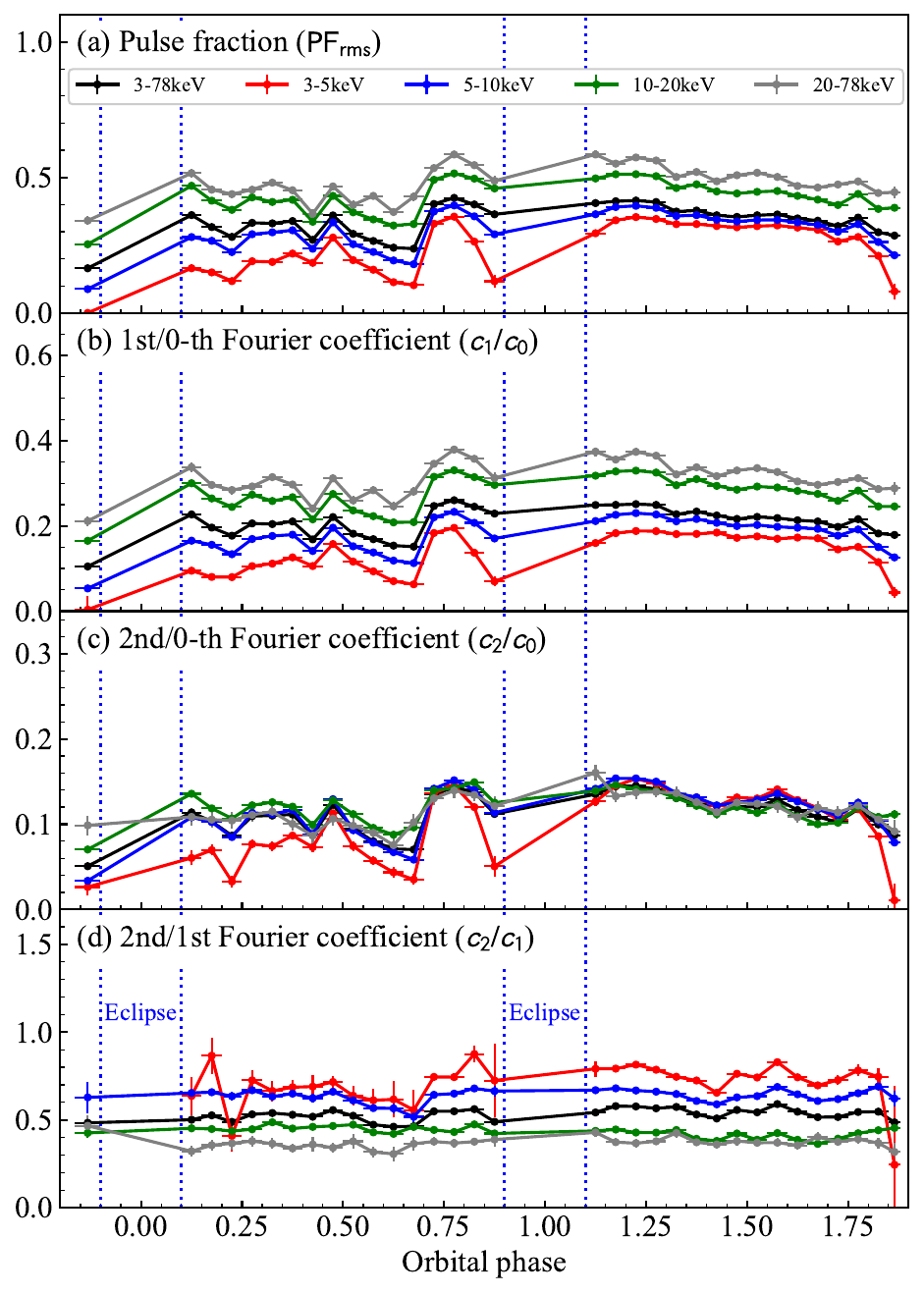}
\caption{\red{Variations of pulse profile parameters along the orbital phase calculated for multiple energy bands. Eclipse phases were excluded because of no pulsed emission.}
}
\label{fig:pulse_profile_parameters}
\end{figure}

In addition to the spectral analysis, we investigated the variability of the pulse profile throughout the orbital phase.
We segmented the observation data into four energy bands: 3--5, 5--10, 10--20, and 20--78 keV, and monitored the variations of the pulse profiles as a function of the orbital phase.
We calculated the pulse fraction for the pulse profiles at various orbital phases, with the definition of the pulse fraction provided in equation (9) in \citetalias{Tamba2023}.
\red{Subsequently, we calculated the ratios of 0-th, 1st, and 2nd harmonics for each energy- and orbital-phase-resolved pulse profiles.
The mathematical definitions of the harmonics are provided in equations (10)-(12) in \citetalias{Tamba2023}.
The calculated pulse fractions and harmonic ratios are presented in Figure \ref{fig:pulse_profile_parameters}.
Among the four energy bands, the lowest, 3--5 keV, exhibits significantly more pronounced variability compared to the other three bands.}

\begin{figure*}[htb!]
\centering
\includegraphics[width=\linewidth]{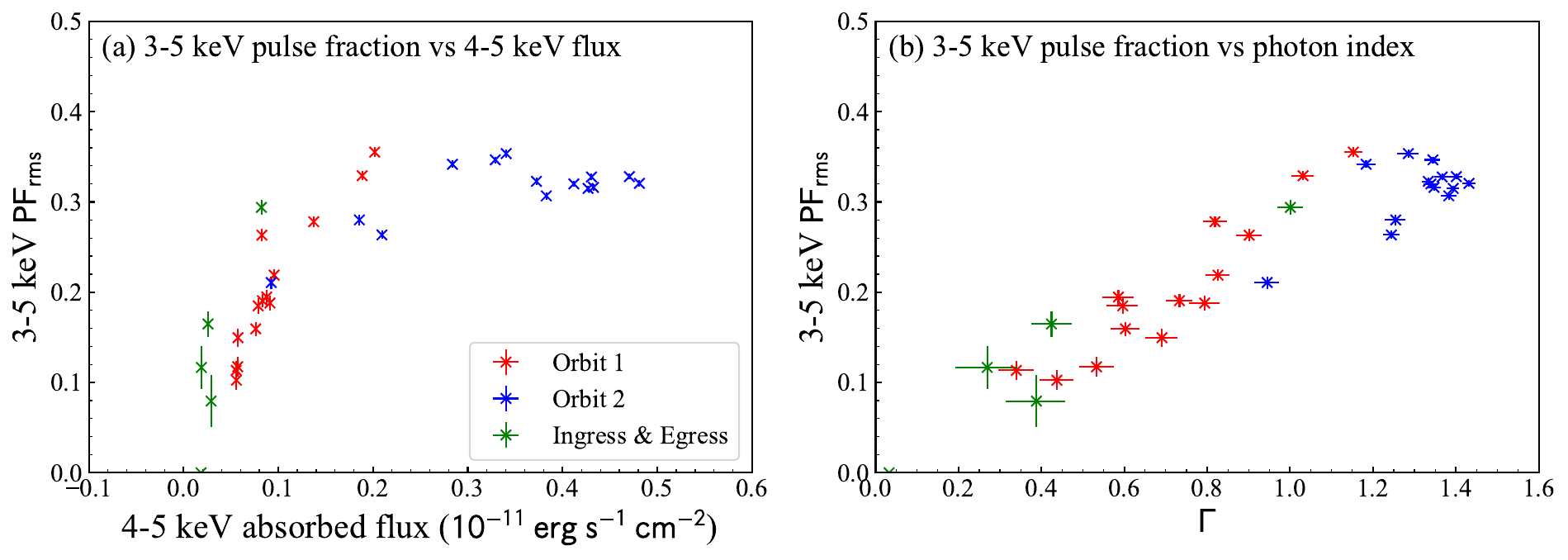}
\caption{3--5 keV pulse fraction plotted as a function of (a) 4--5 keV absorbed flux and (b) photon index.
Different colors represent different types of orbital phases.
}
\label{fig:pulse_fraction_to_continuum}
\end{figure*}

In Figure \ref{fig:pulse_fraction_to_continuum}, we illustrate the correlation between the pulse fraction in the 3--5 keV band and two continuum spectral parameters.
The first parameter is the continuum flux, specifically focused on the 4--5 keV range, which demonstrates a clear positive correlation with the pulse fraction.
The second parameter is the photon index, which also exhibits a positive correlation with the pulse fraction.
It is surprising that the pulse fraction has such a correlation with the continuum spectral variability because the results in Section \ref{subsec:doubly_spectroscopy} suggest that the orbital- and spin-phase variabilities are independent.
The most plausible explanation that reconciles these two results is the presence of an additional non-pulsed emission component that exhibits less variability than the neutron star emission.
In other words, during the low-flux phases, this non-pulsed component becomes more dominant due to its relatively lower variability compared to the pulsed component, resulting in lower pulse fractions.
This additional component is likely to be predominant in the soft X-rays rather than the hard X-rays, as the hard X-ray band does not show significant variability in the pulse fraction.

%% file: section5.tex
\section{Physically-motivated spectral model}\label{sec:advanced_model}

In Sections \ref{sec:spectral_variability} and \ref{sec:spin_phase_resolved}, we have conducted spectral analysis using a phenomenological model.
The variable continuum spectrum can be explained by the Fermi-Dirac cut-off power law, with the photon index being the key parameter characterizing the orbital-phase variability (Section \ref{subsec:phenomeno_spectroscopy}).
The spin-phase-resolved analysis revealed that the spectral variability originates from external factors, distinct from those associated with the neutron star (Section \ref{subsec:doubly_spectroscopy}).
The likely candidate is obscuration by the inhomogeneous stellar wind, as previously suggested in \citetalias{Tamba2023}.
We must provide an explanation for the results of the pulse profile analysis, which indicated the existence of an additional component.
This additional component is likely non-pulsating, dominant in the soft X-ray band, and exhibits less variability compared to the neutron star emission (Section \ref{subsec:pulse_profile}).
The thermal emission from the accretion disk is a strong candidate for this, as it lacks pulsations, typically peaks at $<1\;{\rm keV}$, and is less affected by obscuration from the stellar wind due to its much larger size in comparison to the neutron star.
\red{Reflection by the stellar wind is another possible explanation for the non-pulsating component, but this can be ruled out as the non-pulsating component is observed only in the soft X-ray band.}

We constructed a physically-motivated spectral model to test the hypothesis described above.
This model assumes:
\begin{enumerate}
\renewcommand{\labelenumi}{(\arabic{enumi})}
\item Stable emission from the neutron star represented by the best-fit phenomenological model during the brightest orbital phase ($f_{\rm NS}$, $\Phi=1.40$--$1.45$),
\item Obscuration by the stellar wind, \red{expressed by the photoelectric absorption and Compton scattering due to the stellar wind},
\item Blackbody emission from the accretion disk ($f_{\rm bb}$).
\end{enumerate}
The spectral model can be expressed as
\begin{eqnarray}
\frac{dN}{dE}&=&\exp\left[-\left(\nhim+N_{\rm H}^{\rm PE}\right)\sigma_{\rm abs}(E)\right]\nonumber\\
&&\times\exp\left[-N_{\rm H}^{\rm CS}\sigma_{\rm CS}(E)\right]\times \left(f_{\rm NS}+f_{\rm gauss}\right)\nonumber\\
&&+\exp\left(-\nhim\sigma_{\rm abs}(E)\right)f_{\rm bb}.
\label{eq:attenuation_model_with_bb}
\end{eqnarray}
Here, `PE' and `CS' represent photoelectric absorption and Compton scattering, respectively.
We did not assume any absorption on the blackbody component because it strongly couples with the blackbody normalization when we analyze the NuSTAR energy band.
\red{Therefore, it should be noted that the normalization of the blackbody component (expressed as luminosity in this case) does not reflect the actual change in luminosity but rather the attenuation caused by absorption.}

\begin{figure*}[htb!]
\centering
\includegraphics[width=\linewidth]{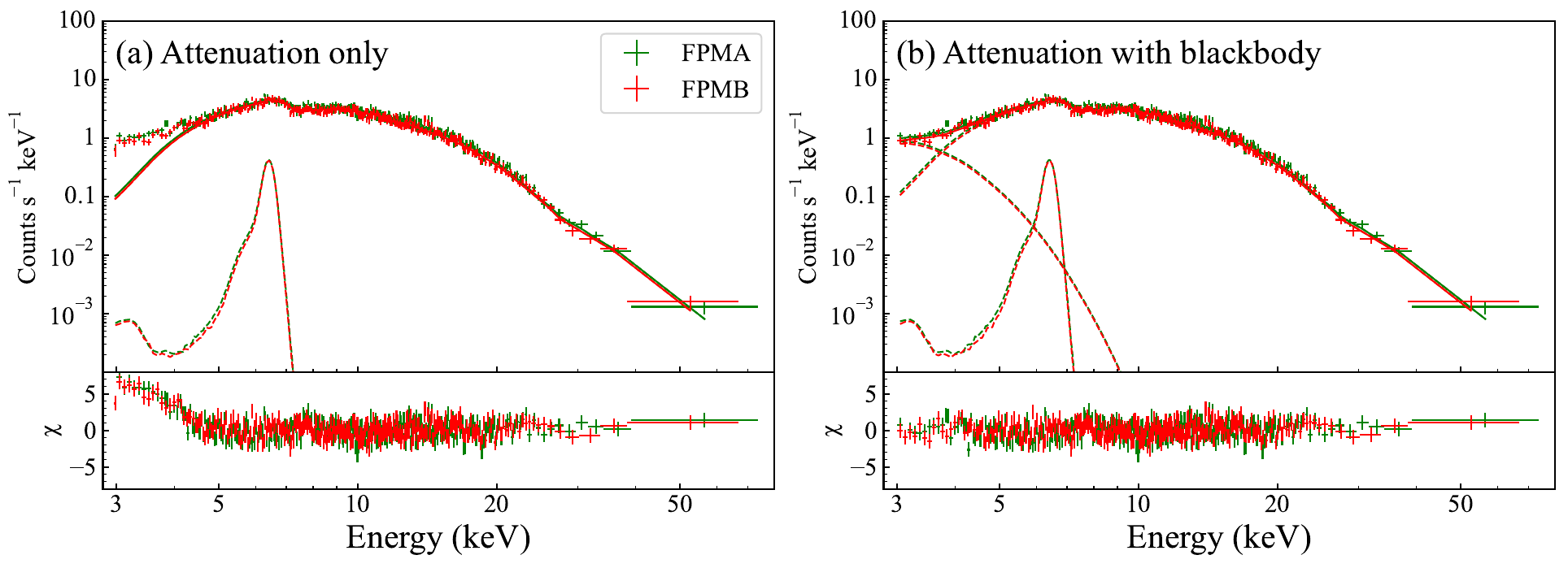}
\caption{
Example of spectral fitting based on physically-motivated models for the orbital-phase-resolved spectrum of $\Phi=0.195-0.120$. (a) Only the attenuation due to the photoelectric absorption and Compton scattering are considered. Fitting was applied to 6--78 keV spectrum. (b) Additional blackbody component with $kT=0.47\;{\rm keV}$ was included in the fitting. Fitting was applied to 3--78 keV spectrum.
}
\label{fig:fitting_example}
\end{figure*}

\begin{figure*}[htb!]
\centering
\includegraphics[width=\linewidth]{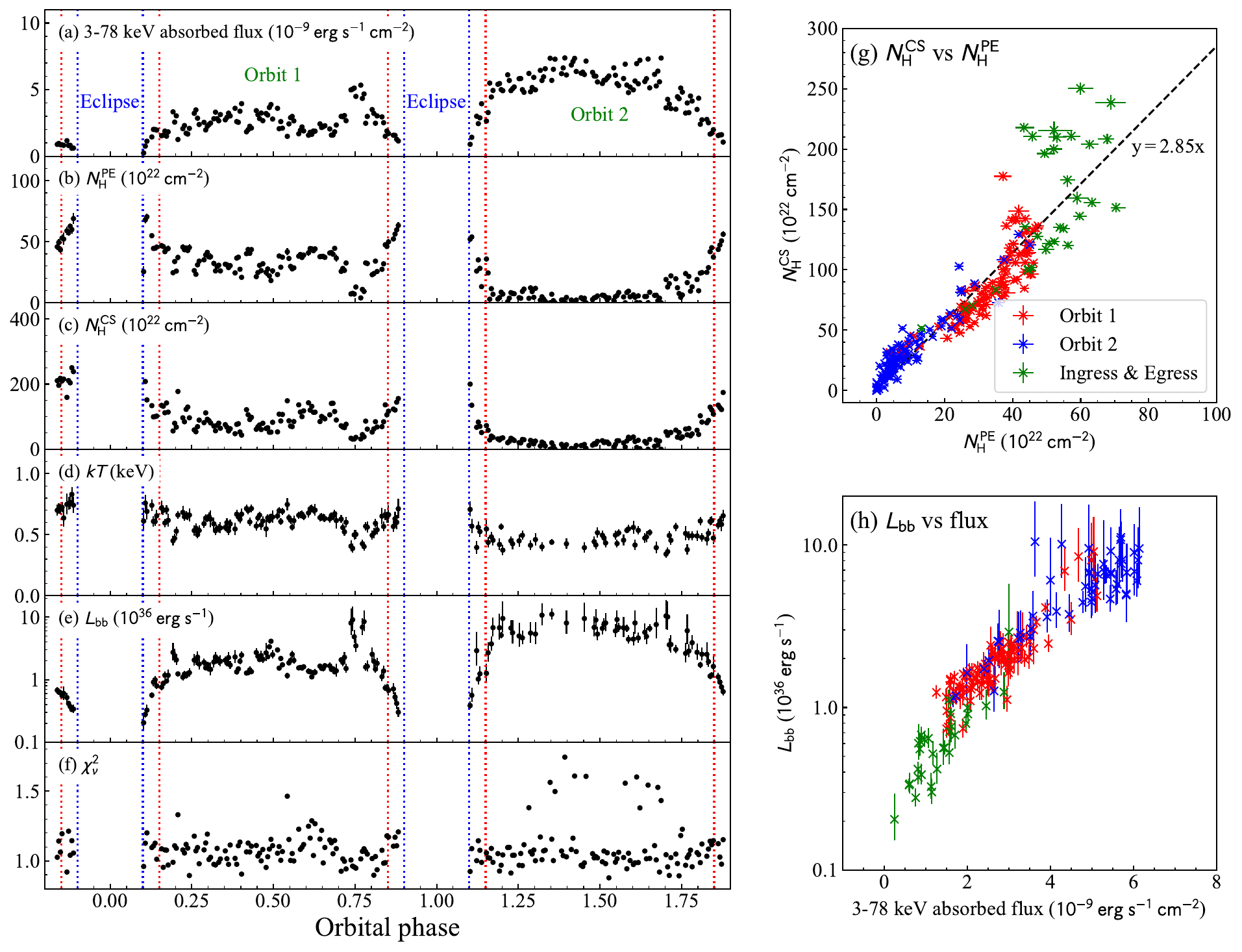}
\caption{(left) Fitting results of orbital-phase-resolved spectra with the physically-motivated spectral model given by equation \eqref{eq:attenuation_model_with_bb}.
$kT$ and $L_{\rm bb}$ are plotted only for the orbital-phase intervals where the blackbody component is significantly detected.
Eclipse phases are indicated by blue dotted lines, and Ingress \& Egress phases are marked with red dotted lines.
(right) Correlation between $\nhpe$ and $\nhcs$ (top), $L_{\rm bb}$ and flux (bottom).
Different colors represent different types of orbital phases.
Error bars represent 90\% confidence levels.
}
\label{fig:orb_physical_fitting}
\end{figure*}

We performed spectral fitting on the orbital-phase-resolved spectra using 3--78 keV photons, employing equation \eqref{eq:attenuation_model_with_bb}.
An example of spectral fitting is presented in Figure \ref{fig:fitting_example}, in which we applied a two-step spectral fitting.
\red{As a first step, only the values of two spectral parameters associated with absorption, $N_{\rm H}^{\rm PE}$ and $N_{\rm H}^{\rm CS}$, were determined by fitting the 6--78 keV spectra (Figure \ref{fig:fitting_example} left).}
Subsequently, we sought the best-fit values for \red{the remaining spectral parameters,} $kT$ and $L_{\rm bb}$, using 3--78 keV spectra (Figure \ref{fig:fitting_example} right).
The left panel of Figure \ref{fig:orb_physical_fitting} presents the best-fit spectral parameters at each orbital phase, with $kT$ representing the blackbody temperature and $L_{\rm bb}$ signifying the blackbody luminosity.
In certain bright phases, notably during Orbit 2, the blackbody component is not significantly detected due to the dominance of the neutron star emission.
As a result, we have excluded these orbital phases from the $kT$ and $L_{\rm bb}$ plot, but they have been retained for other spectral parameters.
\red{Several orbital intervals in Orbit 2 still show large residuals compared to the other intervals, which is also due to the same reason.}
The average value of the reduced chi-squares yields an acceptable value of 1.08, and all the spectra were successfully reproduced by our assumed model.
The flux variability arises from attenuation caused by both photoelectric absorption and Compton scattering.
At its maximum, the column densities for these two interactions reach $N_{\rm H}^{\rm PE}\sim7\times10^{23}\pcmcm$ and $N_{\rm H}^{\rm CS}\sim2.5\times10^{24}\pcmcm$, respectively.
The variations in these two column densities are largely synchronized, with more pronounced attenuation in Orbit 1 compared to Orbit 2, and even more so during the ingress and egress phases.
\red{The blackbody temperature remains relatively stable around $kT\sim0.5\;{\rm keV}$, although there is a systematic variation between the two orbits, with Orbit 1 exhibiting higher temperatures.}
The blackbody luminosity ranges from $L_{\rm bb}\sim10^{36}\;{\rm erg\;s^{-1}}$ to $L_{\rm bb}\sim10^{37}\;{\rm erg\;s^{-1}}$.
It is important to note that the variability in $L_{\rm bb}$ does not necessarily reflect the actual luminosity of the accretion disk as we did not apply any absorption model to account for the blackbody component. 
Instead, the variability in $L_{\rm bb}$ appears to be synchronized with the continuum flux variability, suggesting that the blackbody emission is also subject to attenuation by the stellar wind absorption to some extent.

% 2.85
% bb norm -> absorption
% 

% chatgpt done 2023/10/2

%%%%%%%%%%%%%%%%%%%%%%%%%%%%%%%%%%%%%%%%%%%%%%%%%%%%%%%%%%%%%%%%%%%%%%%%%%%%%%%%%%%%%%%%
%   comments 
%%%%%%%%%%%%%%%%%%%%%%%%%%%%%%%%%%%%%%%%%%%%%%%%%%%%%%%%%%%%%%%%%%%%%%%%%%%%%%%%%%%%%%%%

\begin{comment}

Fig. 5にNHCS vs NHPE, Lbb vs fluxを追加するのが良い？

\end{comment}

%% file: section6.tex
\section{Discussion}
\label{sec:discussion}

In Section \ref{sec:advanced_model}, we successfully explained the spectral variability of Cen~X-3 using a simplified physics-based model.
It suggests that the flux variability is caused by attenuation due to the inhomogeneous stellar wind, and the apparent variability in the photon index arises from different attenuation rates affecting the two emission components: the neutron star and the accretion disk.
The intrinsic variability of the accretion stream does not significantly contribute to the flux variability.
In this section, we assess the physical validity of the analysis results and discuss the emission geometry of Cen~X-3.
%The peculiar photon index variation yielded in the phenomenological spectral analysis (Sections \ref{sec:spectral_variability} and \ref{sec:spin_phase_resolved}) has eventually been compensated by the thermal emission from the accretion disk.

\subsection{Stellar wind}

There are four parameters that describe the continuum spectra, with only $\nhpe$ and $\nhcs$ contributing to the broadband spectral variability.
Although they are fitted independently, the two column densities exhibit a strong correlation throughout the observation, as depicted in Figure \ref{fig:orb_physical_fitting}(g).
The correlation can be approximated as $\nhcs\sim2.85\nhpe$.
The constant ratio between $\nhpe$ and $\nhcs$ suggests that the inhomogeneous stellar wind obscuring the neutron star maintains a consistent ionization degree regardless of the orbital phase.
This finding supports the presence of a stellar wind with an uneven geometric distribution but uniform ion fraction.
A plausible scenario is a clumpy stellar wind distributed on the line of sight.
Assuming the orbital velocity of the neutron star is $v=4.4\times10^{7}\;{\rm cm\;s^{-1}}$ and a typical variation timescale of $\Delta\Phi=0.01$, we can calculate the typical size of a stellar wind clump as follows:
\begin{eqnarray}
R_{\rm c}=v\porb \Delta\Phi=8\times10^{10}\left(\frac{\Delta\Phi}{0.01}\right)\;{\rm cm}.
\label{eq:clump_size}
\end{eqnarray}
We can also estimate the typical hydrogen number density of the clumps as 
\begin{eqnarray}
n_{\rm c}=\frac{N_{\rm H}}{R_{\rm c}}=1\times10^{13}\left(\frac{N_{\rm H}}{10^{24}\;{\rm cm^{-2}}}\right)\left(\frac{\Delta\Phi}{0.01}\right)^{-1}\;{\rm cm^{-3}}.
\end{eqnarray}
These values are consistent with the ones obtained in \citetalias{Tamba2023} ($R_{\rm c}\sim9\times10^{10}\;{\rm cm}$ and $n_{\rm c}\sim3\times10^{12}\;{\rm cm^{-3}}$).

The analysis in this paper strongly suggests that the spectral variability of Cen~X-3 is primarily caused by the inhomogeneous stellar wind, a hypothesis supported by many previous studies \citep[e.g.,][]{Tsunemi1996, Naik2011, Liu2024}.
The contribution from variations in the accretion rate is smeared out in the two orbital cycles, which is estimated to cause $\sim20\%$ flux variability in \citetalias{Tamba2023}.
However, there are certain aspects that our hypothesis does not fully explain.
The most notable is the systematic spectral difference between Orbit 1 and Orbit 2; Orbit 2 consistently exhibits higher fluxes than Orbit 1.
This difference could stem from variations in the distribution of the stellar wind or possibly an intrinsic transition from a low state to a high state \citep{Raichur2008}.
Further investigation is warranted, and more sophisticated observations, such as precise measurements of the Fe line from the stellar wind using XRISM \citep{Tashiro2020}, may shed light on this matter.

\subsection{Accretion disk}

As depicted in Figure \ref{fig:orb_physical_fitting}(d), the temperature of the blackbody emission remains around $kT\sim0.5\;{\rm keV}$ throughout the observation, likely indicating the temperature at the innermost radius of the accretion disk.
Utilizing the standard accretion disk model \citep{Shakura1973}, the mid-plane temperature of the accretion disk at a specific radius $r$ can be expressed as
\begin{eqnarray}
T_{\rm c}\sim1.4\times10^4\left(\frac{\dot{M}}{10^{16}\;{\rm g\;s^{-1}}}\right)^{3/10}
\left(\frac{M_{\rm ns}}{M_\odot}\right)^{1/4}\left(\frac{r}{10^{10}\;{\rm cm}}\right)^{-3/4}\;{\rm K}.
\end{eqnarray}
$\dot{M}$, $M_{\rm ns}$, and $r$ represent the accretion rate, the mass of the neutron star, and the radius of the observed point, respectively.
By substituting $\dot{M}=3.2\times10^{17}\;{\rm g\;s^{-1}}$, $M_{\rm ns}=1.34M_{\odot}$, and $r=3\times10^{8}\;{\rm cm}$ (corresponding to the \alfven\ radius of Cen~X-3), we obtain $kT_{\rm c}\sim0.05\;{\rm keV}$.
However, the observed value of $kT\sim0.5\;{\rm keV}$ is an order of magnitude higher than this theoretical prediction.
This difference can be accounted for by considering irradiation from the neutron star.
If the accretion disk absorbed all the emission from the neutron star, the temperature would approach the Compton temperature \citep{Done2010}.
This temperature represents the equilibrium point where the processes of Compton up-scattering and down-scattering are balanced.
For Cen~X-3, this temperature is calculated to be $\Theta=2.6\;{\rm keV}$, suggesting that a portion of the neutron star emission contributes to heating the accretion disk.
%This temperature is expressed by $\Theta=\int N(\epsilon)\epsilon^2d\epsilon/4\int N(\epsilon)\epsilon d\epsilon$, which becomes $\Theta=2.6\;{\rm keV}$ for Cen~X-3 spectrum.
%The measured $kT\sim0.5\;{\rm keV}$ lies between $kT_{\rm c}$ and $\Theta=2.6\;{\rm keV}$, which means a part of the neutron star emission contributes to the heating of the accretion disk.

Figure \ref{fig:orb_physical_fitting}(e) illustrates the variation in the accretion disk luminosity $L_{\rm bb}$, ranging from $10^{36}\;{\rm erg\;s^{-1}}$ to $10^{37}\;{\rm erg\;s^{-1}}$.
As shown in Figure \ref{fig:orb_physical_fitting}(h), $L_{\rm bb}$ has a strong correlation with the continuum flux.
However, as previously highlighted in Section \ref{sec:advanced_model}, this variability in luminosity may not accurately reflect the actual changes in the accretion disk luminosity as this parameter strongly couples with the absorption.
In reality, there is no physical mechanism that could cause such rapid changes in accretion disk luminosity.
Therefore, it is reasonable to attribute this luminosity variability to partial absorption by clumps in the stellar wind.
Assuming a blackbody temperature of $kT=0.5\;{\rm keV}$, a decrease in luminosity by an order of magnitude could be explained by absorption column densities of $\nhpe=1\times10^{23}\pcmcm$ and $\nhcs=2.85\times10^{23}\pcmcm$.
These values are significantly smaller than the maximum values of $\nhpe$ and $\nhcs$ seen in Figure \ref{fig:orb_physical_fitting}.
Consequently, the accretion disk experiences partial absorption by the stellar wind, leading to effective column densities that are much smaller than those of the neutron star emission.
This, in turn, results in the apparent spectral hardening and reduced pulse fraction during the low-flux phases.

The characteristic timescale of \red{relative increase and decrease in} $L_{\rm bb}$ provides further evidence for the existence of the accretion disk emission.
During the ingress and egress phases, the disk emission diminishes and re-emerges, respectively, over a timescale of $\Delta\Phi\sim0.05$, which corresponds to $\Delta t\sim9\;{\rm ks}$.
Assuming the binary separation of $R_{\rm sp}=17.9R_{\odot}$ \citep{Sanjurjo2021}, the typical size of the blackbody emission region can be estimated to be $\sim4\times10^{11}\;{\rm cm}$.
The size of the accretion disk can be estimated based on the size of the Roche lobe \citep{Paczynski1971}, given by
\begin{eqnarray}
R_{\rm L}=0.462\left(\frac{M_{\rm ns}}{M_{*}}\right)^{\frac{1}{3}}\sim2.3\times10^{11}\;{\rm cm},
\end{eqnarray}
where we substitute the masses of the neutron star and the companion star, $M_{\rm ns}=1.34M_{\odot}$ and $M_{*}=20.5M_{\odot}$, respectively.
The sizes of the blackbody emission region and the estimated accretion disk are comparable, which supports the validity of the assumed spectral model.
Furthermore, it confirms that the size of the accretion disk is much larger than the typical size of the stellar wind clumps (equation \ref{eq:clump_size}).

%chatgpt done 2023/10/2

%%%%%%%%%%%%%%%%%%%%%%%%%%%%%%%%%%%%%%%%%%%%%%%%%%%%%%%%%%%%%%%%%%%%%%%%%%%%%%%%%%%%%%%%
%   comments 
%%%%%%%%%%%%%%%%%%%%%%%%%%%%%%%%%%%%%%%%%%%%%%%%%%%%%%%%%%%%%%%%%%%%%%%%%%%%%%%%%%%%%%%%

\begin{comment}

\end{comment}

%% file: section7.tex
\section{Conclusions}\label{sec:conclusions}

We conducted a NuSTAR observation on Cen~X-3 spanning two consecutive orbital cycles to investigate its spectral variability.
The light curves showed more variations in lower energy bands compared to higher energy bands.
These variations were characterized by changes in the photon index, with a softer photon index observed during higher-flux phases and a harder one during lower-flux phases.
While the variability of the spectral parameters did not exhibit any correlation with the spin phase, the pulse fraction displayed a strong correlation with these spectral parameters, increasing during higher-flux phases and decreasing during lower-flux phases.
By combining these results, we have arrived at a conclusion regarding the physical origin of the spectral variability.
The intrinsic emission from the neutron star remains stable, and the spectral variability primarily arises from attenuation by non-uniformly distributed clumps in the stellar wind.
The apparent variations in the photon index and pulse fraction are attributed to an additional non-pulsed emission component, specifically thermal emission from the accretion disk.
We estimate that the stellar wind clumps have a typical size of $R_{\rm c}\sim8\times10^{10}\;{\rm cm}$ and a typical hydrogen number density of $n_{\rm c}\sim1\times10^{13}\;{\rm cm^{-3}}$.
The thermal emission from the accretion disk can be modeled as a blackbody with a temperature of $kT\sim0.5\;{\rm keV}$.
The maximum luminosity of the blackbody emission is $\sim10^{37}\;{\rm erg\;s^{-1}}$, and it experiences some attenuation due to partial absorption by the clumpy stellar wind.

%chatgpt done 2023/10/2

%%%%%%%%%%%%%%%%%%%%%%%%%%%%%%%%%%%%%%%%%%%%%%%%%%%%%%%%%%%%%%%%%%%%%%%%%%%%%%%%%%%%%%%%
%   comments 
%%%%%%%%%%%%%%%%%%%%%%%%%%%%%%%%%%%%%%%%%%%%%%%%%%%%%%%%%%%%%%%%%%%%%%%%%%%%%%%%%%%%%%%%

\begin{comment}

\end{comment}